\documentclass[floatfix,superscriptaddress,amsmath,showpacs,showkeys,aps,notitlepage,prb]{revtex4-1}

\usepackage{t1enc}
\usepackage{epsfig}
\usepackage{bm}
\usepackage[normalem]{ulem}
\usepackage{times}
\usepackage{verbatim}
\usepackage{amssymb}
\usepackage{xcolor}

\begin{document}

\title{Topological phase diagrams of the frustrated Ising ferromagnet}

\author{Alejandro Mendoza-Coto}%
\email{alejandro.mendoza@ufsc.br}
\affiliation{Departamento de F\'\i sica, Universidade Federal de Santa Catarina, 88040-900 Florian\'opolis, Brazil}%

\author{Danilo Emanuel Barreto de Oliveira}%
\affiliation{Departamento de F\'\i sica, Universidade Federal de Rio Grande do Sul, 91501-970 Porto Alegre, Brazil}%

\author{Lucas Nicolao}%
\affiliation{Departamento de F\'\i sica, Universidade Federal de Santa Catarina, 88040-900 Florian\'opolis, Brazil}%

\author{Rogelio D\'iaz-M\'endez}%
\affiliation{Department of Physics, KTH Royal Institute of Technology, SE-10691 Stockholm, Sweden}%

\date{\today}

\keywords{Topological phases, Frustrated ferromagnet, Reentrant behavior}

\begin{abstract}
The emergence of complex modulated structures in the magnetization pattern of thin films is a well-established experimental phenomenology caused by the frustrating effects of competing interactions.
Using a coarse-grained version of the Ising ferromagnet with dipolar interactions, we develop a method that use the information from the microscopic Hamiltonian to predict the specific topological phases present in the temperature-external magnetic field phase diagram.
This is done by the combination of mean-field variational calculations and the Renormalization Group equations from the classical theory of two-dimensional melting.
In this framework, we are able to distinguish when the orientational and translational symmetries are broken, discriminating between the ordered and disordered states of the system for all temperatures and fields. 
We observe that the reentrance developed by the $H$-$T$ phase diagrams in the regime of weak dipolar interactions is directly related with the appearance of anomalous topological transitions. 
These results motivate the realization of new experiments on magnetic thin films in order to explore the topological properties of the magnetic textures, allowing to identify new exotic phases in these materials.
\end{abstract}

\maketitle


\section{Introduction}

The complex patterned phases appearing in two dimensional systems with competing interaction have been extensively studied in the literature~\cite{BoMaWhDe1995,So1987,AnBrJo1987,AlStBi1990,CaCaBiSt2011,PiBiStCa2010,DeSa1996}. 
The interest in such systems goes beyond the technological relevance.
Since in two dimensions the fluctuations can be particularly strong even at very low temperature, systems are usually unable to stabilize crystalline patterns and consequently only weaker forms of order are observed~\cite{MeWa1966,Ne1978,KoTh1973,ToNe1981}. 
Example of these phases are the nematic phase in two dimensional systems of stripes, and the 2D solid and hexatic phases in two dimensional systems of particles~\cite{NeHa1979,ToNe1981,AbKaPoSa1995}. 
Theoretically speaking, the melting of these phases, meaning the restoration of one or more symmetries of the system, is expected to occur through some topological mechanism
with increasing temperature~\cite{Ko2016}. 

A model system presenting this phenomenology is the so-called isotropic dipolar ferromagnet with strong perpendicular anisotropy~\cite{GaDo1982,DeMaWh2000,DeMaBoWh1997,PoVaPe2003}. In this model the competition between the ferromagnetic interaction and the dipolar antiferromagnetic interaction produces nontrivial magnetic textures of the out-of-plane magnetization.
This model has been successfully used to study various thermodynamic and magnetic properties of ultrathin ferromagnetic films with perpendicular anisotropy of Fe/Cu(001) \cite{SaLiPo2010, SaRaViPe2010, Sa2016}, including the recently observed inverse melting transitions in the presence of applied magnetic field \cite{MeNiDi2019, MeBiCaSt2016, MeSt2012}.
A commonly observed structure in these magnetic systems is the stripe configuration, in which the sign of the local magnetization oscillates when moving along a certain direction in the system.
This stripe phase posses only local positional order, though a weak orientational order is stable with power-law decaying correlations. 
The latter is usually called the nematic phase. 
Upon increasing temperature this nematic phase melts into a liquid of stripes with only local positional and orientational order.
This is precisely the stripe melting scenario \cite{ToNe1981,AbKaPoSa1995} predicted by the Kosterlitz-Thouless-Halpering-Nelson-Young (KTHNY) classical theory of two-dimensional melting \cite{KoTh1973,Ne1978,HaNe1978,NeHa1979,Yo1979,St1988,Ko2016}. This theory describes a class of phase transitions produced by the proliferation of topological defects, such as vortices, disclinations and dislocations. The low temperature phases of these systems, characterized by a non-zero density of such defects, are commonly called topological phases.

Although several works \cite{ToNe1981,AbKaPoSa1995,CaMiStTa2006,BaSt2007,NiSt2007,BaMeSt2013, MeStNi2015} have considered the properties of these transitions, fewer attempts have been made to characterize the properties of the nematic transition when an external magnetic field is applied. 
The external field naturally favors magnetic domains oriented along its direction, producing an enlargement of the modulation length and an asymmetry of the positive and negative domains in the stripe pattern. 
These effects are very hard to relate with the topological properties of the stripe phase on physical grounds, which explain the lack of literature on the detailed characterization of the topological phases within the stripe region. 

A similar situation is present regarding the topological properties of the bubble phase observed for larger fields.
Although experimentally \cite{SaLiPo2010, SaRaViPe2010, Sa2016} and computationally accessible \cite{DiMu2010,CaCaBiSt2011,MeNiDi2019}, a topological characterization of the bubble lattice varying temperature and magnetic field is still lacking. 
While it can be argued that perfect bubble configurations share the same symmetries of a triangular lattice of particles \cite{HaNe1978,NeHa1979,Yo1979}, it is not clear whether the elastic effective Hamiltonian of the bubble patterns coincide with that of a triangular crystal.
Consequently, the topological phases and its behavior within the bubble region of the $H$-$T$ phase diagram, although possibly related somehow to the problem of the two dimensional melting of triangular crystals, lack a theoretical description. 

Another intriguing issue is the effect on the topological phases produced by the presence of reentrance in the phase diagrams like the one observed experimentally in ultrathin ferromagnetic films with perpendicular anisotropy \cite{SaLiPo2010, SaRaViPe2010, Sa2016}. 
These magnetic systems can present strong reentrances in the $H$-$T$ plane, depending on the relative strength of the dipolar and the ferromagnetic interactions. 
For systems with weak enough repulsive interaction the $H$-$T$ phase diagram develops an inverse melting in a certain range of the external field~\cite{MeNiDi2019}. 
This anomaly in the phase diagram is accompanied by an anomalous behavior of the Young modulus of the patterns \cite{PoGoSaBiPeVi2010, MeSt2012}, which suggests possible modifications in the topological order of the magnetic texture.


In this work we begin defining the microscopic Hamiltonian for the Ising-dipolar model and, after a coarse-graining procedure, we obtain a mean-field free energy functional of the local magnetization. This allows us to find the local form of the magnetic modulated pattern, which in turn is used to calculate the bare elastic stiffnesses of the stripes and bubbles configurations.
At this point we take advantage of the classical theory of two-dimensional melting, and use the elastic stiffnesses previously calculated as an input for the well established Renormalization Group (RG) equations of the theory. These equations basically allows for a calculus of the renormalized stiffnesses due to the presence of topological defects \cite{HaNe1978,NeHa1979,Yo1979,ToNe1981}. In this context, the phase with short-range order of certain type is identified by the values of $H$ and $T$ where the appropriate renormalized stiffness goes to zero. For example, the stripes nematic phase can be identified as the region where the renormalized orientational stiffness of the stripes pattern is different from zero \cite{ToNe1981}.
In this way, we are able to identify within the stripes and bubbles regions in the mean-field phase diagrams the regions corresponding to the {\it stripes nematic} and {\it stripes liquid} phases, as well as the {\it 2D solid}, {\it hexatic} and {\it liquid} bubble phases. 
Such information allows to study both the effect of the external field in the behavior of topological phases and the role of the reentrance of the $H$-$T$ phase diagrams in the appearance of anomalous topological transitions. 

This paper is organized as follows. 
In sections two and three the microscopic model is described and the mean-field free energy functional is built, to be used as starting point in obtaining the bare and renormalized elastic stiffnesses of the different phases. 
In section four we discuss the numerical results on the topological phase diagrams varying the relative strength of the dipolar interaction.
Section five focuses on the characterization of some critical properties of the topological transitions between the different phases. 
The general conclusions are discussed in section six.

\section{Model}
We begin considering a 2D Ising spin model with nearest-neighbor ferromagnetic coupling $J$ and a long-range dipolar interaction of
relative strength $\delta^{-1}$. This model can be described by the Hamiltonian~\cite{PiCa2007,DiMu2010}
\begin{equation}
\label{Hmic}
 \mathcal{H}=-\frac{J}{2}\sum_{\langle i,j\rangle}s_is_j+\frac{J}{2\delta}\sum_{i\neq j}
 \frac{s_is_j}{\vert \vec{x}_i-\vec{x}_j\vert^3}
 -h\sum_is_i
 \mathrm{,}
\end{equation}
where $h$ represents the external magnetic field applied perpendicular to the plane of the system.
In the Fourier space the previous expression can be rewritten as 
\begin{equation}
  \mathcal{H}=\frac{1}{2}\int_{BZ}\frac{d^2\vec{k}}{(2\pi)^2}\hat{A}(k)\hat{s}(\vec{k})\hat{s}(-\vec{k})-\int_{BZ}\frac{d^2k}{(2\pi)^2}h(k)\hat{s}(-k),
  \label{Hmick}
\end{equation}
where the function $\hat{A}(k)$ represent the so called fluctuation spectrum and the subindex ``$BZ$'' stands for the first Brilloin zone of the 
original square lattice. The fluctuation spectrum in the long wave limit $k\rightarrow0$, as obtained by Cannas et. al.~\cite{PiCa2007}  will be given approximately by the 
expression
\begin{align}
  \begin{split}
 \hat{A}(k) =&-2J\left(2-\frac{k^2}{2}\right)+\frac{J}{\delta}\left(\frac{2\pi^2}{3}+2\zeta(3)-2\pi k+k^2\right)\\
 =&-\frac{\left(6\zeta(3)-\pi^2+2\delta(\pi^2+3\zeta(3)-6)-12\delta^2\right)}{3\delta(1+\delta)}J+\frac{1+\delta}{\delta}J\left(k-\frac{\pi}{1+\delta}\right)^2.
  \end{split}
  \label{Ak1}
\end{align}

The spin configurations which minimize the Hamiltonian (\ref{Hmick}) possess a characteristic wave-vector which is lower\cite{PiCa2007} than the one minimizing the function $\hat{A}(k)$ in eq. (\ref{Ak1}), given by $k_0=\pi/(1+\delta)$. This means that in the regions of large $\delta$ the modulation in $s_i$ spans many lattice sites. In this limit, a coarse grained description is fully justified and the resulting effective Hamiltonian is:
\begin{equation}
\mathcal{F}[\phi]=\frac{1}{2}\int\frac{d^2\vec{k}}{(2\pi)^2}\hat{A}(k)\hat{\phi}(\vec{k})\hat{\phi}(-\vec{k})
+k_BT\int d^2x\ S(\phi)-h\int d^2\vec{x}\ \phi(\vec{x}),
\label{Fener}
\end{equation}
This form of the effective Hamiltonian coincides with the mean-field free energy of eq. (\ref{Hmic}) in the continuum limit \cite{MeBiCaSt2016,MeNiDi2019}.
Here the scalar-field order parameter $\phi(\vec{x})$ is the local magnetization, and $S(\phi)=\frac{(1+\phi)}{2}\log\left(\frac{(1+\phi)}{2}\right)+\frac{(1-\phi)}{2}\log\left(\frac{(1-\phi)}{2}\right)$ represents the local mean-field entropy of the model given by eq. (\ref{Hmic}). 
Now we proceed with the construction of the dimensionless variational Hamiltonian, appropriate for the numerical analysis. 
The rescaling of momenta and lengths are taken in the form $\vec{k}/k_0=\vec{k}'$ and $k_0\vec{x}=\vec{x}'$, which leads to a rescaling of the Fourier transforms as $\hat{\phi}'(\vec{k}')=k_0^2\hat{\phi}(\vec{k}/k_0)$. 
In the new variables, eq. (\ref{Fener}) can be rewritten as:
\begin{equation}
\mathcal{F}[\phi']=\frac{\hat{A}(k_0)}{k_0^2}\left[\frac{1}{2}\int \frac{d^2\vec{k}'}{(2\pi)^2}\hat{A}'(k')\hat{\phi}'(\vec{k}')\hat{\phi}'(-\vec{k}')+
T'\int d^2\vec{x}'S(\phi'(\vec{x}'))-h'\int d^2\vec{x}'\ \phi(\vec{x}')\right],
\end{equation}
where
\begin{align}
A'(\vec{k}')&= -1+a(k-1)^2\\
a&= 3\pi^2/\left(\pi^2-6\zeta(3)+2\delta(6-\pi^2-3\zeta(3))+12\delta^2\right)\\
T'&= k_BT/\hat{A}(k_0)\\
h'&= h/\hat{A}(k_0).
\end{align} 
Finally we choose the units of energy as the quantity $\hat{A}(k_0)/k_0^2$, which leads to the dimensionless Hamiltonian functional\cite{MeBiCaSt2016,MeNiDi2019}
\begin{equation}
\mathcal{F}[\phi] = \frac{1}{2}\int \frac{d^2\vec{k}}{(2\pi)^2}\hat{A}(k)\hat{\phi}(\vec{k})\hat{\phi}(-\vec{k})+
T\int d^2\vec{x}S(\phi(\vec{x}))-h\int d^2\vec{x}\ \phi(\vec{x}),
\label{ffun}
\end{equation}
where prime symbols have been omitted to simplify the notation. It is worth noticing that the above effective Hamiltonian depends only on three parameters: temperature $T$, external magnetic field $H$ and the curvature of the fluctuation spectrum $a$, which is controlled only by the relative strength of the dipolar interaction $\delta$ - the weaker this interaction is, the smaller is $a$.

This kind of mean field free energy have been extensively used as a starting point to construct the phase diagram of systems with competing interactions by looking at the local structure of the magnetization\cite{MeNiDi2019}. At the mean field level a direct minimization of the free energy functional (\ref{ffun}) allows the construction of a phase diagram that contains three different perfectly ordered phases, the {\it stripes phase}, the {\it bubbles phase} and the {\it homogeneous phase}. The so-called stripes phase is defined by a solution in which the local magnetization is given by a one dimensional modulation in the general form:
\begin{equation}
 \phi(\vec{x})=c_0+\sum_{n=1}^{\infty}c_n\cos(nk_0x),
\end{equation}
where the coefficients $c_n$'s represent the Fourier amplitudes of the solution and $k_0$ represent the wave vector characterizing the spatial period of the modulation. Another possible solution of the free energy minimization problem corresponds to the bubbles phase. This solution is given by a two dimensional magnetization pattern in which bubbles magnetized in the opposite direction to the external magnetic field form a triangular array. Mathematically this solution is given by: 
\begin{equation}
\phi(\vec{x})=c_0+\sum_{n=1}^{\infty}c_n\cos(k_0\vec{b}_n\cdot\vec{x}), 
\end{equation}
where again the coefficients $c_n$'s represent the Fourier amplitudes of the solution and $k_0$ represent the wave vector characterizing the spacial period of the triangular lattice of bubbles. Finally, the set of vectors $\vec{b}_n$'s are such that they form a triangular lattice of points with unitary lattice spacing.

The last possible solution of the mean field free energy minimization problem corresponds to the homogeneous solution, which represent a phase in which the average local magnetization is spatially homogeneous. Mathematically, this solution is given by a local magnetization of the form: 
\begin{equation}
\phi(\vec{x})=c_0, 
\end{equation}
where $c_0$ represent the average magnetization. Within the mean-field approximation, this featureless phase is equivalent to a paramagnetic phase.

\begin{figure}[th!]
\begin{center}
\includegraphics[width=0.5\columnwidth]{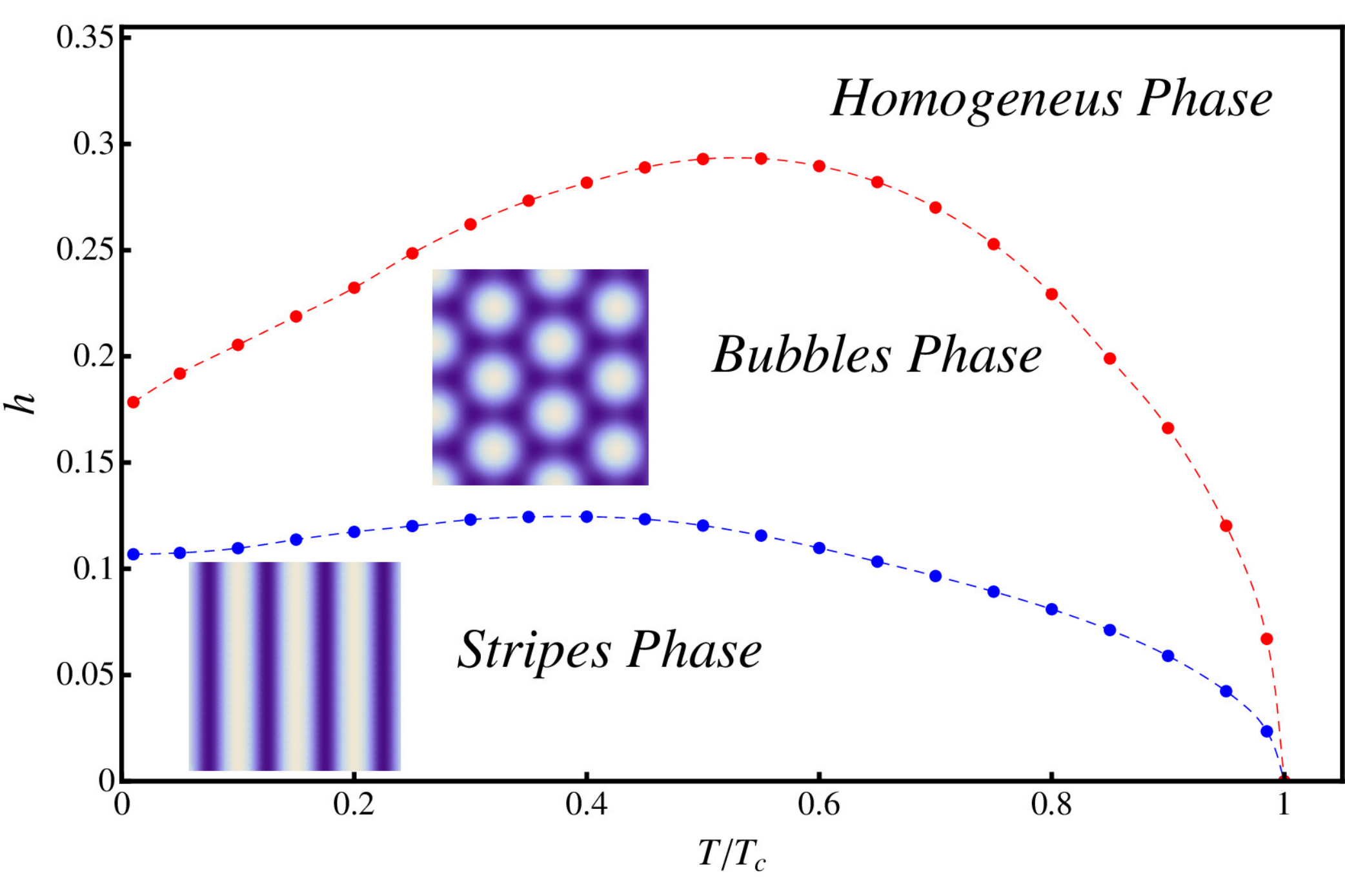}
\end{center}
\caption{
An example of a $H$-$T$ mean field phase diagram for the case $a=0.5$, with illustrations of the stripe and bubbles phases for particular values of $H$ and $T$. }
\label{fig0}
\end{figure}

A comparison of the free energy of each kind of solution, after the minimization procedure, allows the construction of the external field ($h$) versus temperature ($T$) mean field phase diagram. Although the properties of the mean field phase diagram of this system are well understood~\cite{MeBiCaSt2016, MeNiDi2019}, for the sake of completeness we show in Fig.\ref{fig0} the resulting phase diagram for a given value of the fluctuation spectrum curvature parameter ($a=0.5$). The reentrance of the homogeneous phase observed in this phase diagram is an unusual effect due to both the presence weak enough repulsive interactions (low $a$ regime)~\cite{PoGoSaBiPeVi2010, MeBiCaSt2016} and of the large entropy variation due to small variations of the local magnetization close to its saturation value~\cite{MeNiDi2019}.

Although the phase diagram resulting from the mean field approximation allows the identification of phases in which the average local magnetization develops modulated patterns with long range positional and orientational order, in two dimensions it is expected~\cite{ToNe1981,Ne1978,NeHa1979} that these perfectly ordered phases  lose their long range order due to the presence of thermal fluctuations. This means that the stripes and bubbles phases do not exist at any finite temperature beyond the mean field description, although locally the system displays magnetic textures made of stripes and bubbles domains in different regions of the actual phase diagram observed in experiments~\cite{SaLiPo2010, SaRaViPe2010} or in numerical simulations~\cite{MeNiDi2019}. In this scenario, the construction of a phase diagram beyond the mean field level should be made distinguishing each phase by its symmetries. For a two dimensional system in which the local magnetization develops a non-trivial texture constituted by stripes or bubbles, it is natural to identify the different phases by characterizing the behavior of the translational and rotational symmetries of the magnetic pattern. This is developed throughout the next sections within the classical theory of two-dimensional melting.

\section{Topological phases}

In order to find a more accurate description of the phase diagram of our system we must include the role of the low energy excitations over the perfectly ordered patterns minimizing the effective Hamiltonian (\ref{ffun}), already considered in the construction of the mean field phase diagram. The low energy excitations to take into account in our case are of two types: smooth and extended deformations of the reference pattern and strong but localized deformations in the form of topological defects.

As it is well-known from KTHNY theory, these are the most relevant fluctuations in the melting of modulated patterns. This theory concludes that isotropic two-dimensional stripe systems can at best break the rotational symmetry into a nematic phase at low temperatures, characterized by quasi-long-range orientational order and local positional order. Upon heating, this nematic phase will eventually melt into a disordered phase in which both the positional and the orientational order exist only locally.  

On the other hand it is expected, if the bubbles structure behaves as a triangular lattice of particles, that it could in principle melt following a sequence of 2D solid, hexatic and liquid phases as temperature is increased from the low temperature regime. As usual, in the 2D solid phase the system should exhibit quasi-long-range positional order and long-range orientational order; in the hexatic phase, local  positional order and quasi-long-range orientational order; and in the liquid of bubbles local positional and orientational order.         

To study the topological order of the modulated structures we use as starting point the free energy functional of Eq.~(\ref{ffun}) as our effective coarse grained Hamiltonian depending on the local order parameter $\phi(\vec{x})$. The partition function of the effective model can be written as:
\begin{equation}
  \mathcal{Z} = \int \mathcal{D}\phi \,\mathrm{e}^{-\beta\mathcal{F}[\phi]}.
\end{equation}

The evaluation of this partition function implies the consideration of all realizations of $\phi(\vec{x})$. However, it is expected in the low temperature regime that the configurations which contribute the most to the partition function are those in the vicinity of the configuration $\phi_0(\vec{x})$ that minimizes $\mathcal{F}[\phi]$. In this scenario, the role of small fluctuations around $\phi_0(\vec{x})$ are usually treated perturbatively. In our context, the classical theory of two-dimensional melting identifies two kinds of relevant fluctuations. The first group corresponds to small, smooth and continuous deformations of the pattern $\phi_0(\vec{x})$, which can be obtained including a deformation field in the phase of the modulation. The energy cost of these kind of fluctuations leads to effective \emph{elastic} Hamiltonians for each kind of pattern, bubbles or stripes, which are quadratic in the deformation fields and naturally leads to the existence of Golstone modes. The second relevant contribution to the partition function comes from localized singular deformations of the pattern $\phi_0(\vec{x})$, normally called topological defects. 

In the subsequent sections we perform this kind of analysis, and obtain effective elastic Hamiltonians which coincides with those of the KTHNY theory. Consequently, we can take advantage of the well established Renormalization Group (RG) analysis of this theory, identifying the different topological phase transitions occurring for each kind of pattern. This analysis ultimately allows us to calculate the topological properties of the system at any point in the phase diagram.

\subsection{Topological characterization of the stripes phase}

As discussed in the literature~\cite{MeNiDi2019}, one of the possible configurations minimizing a free energy functional of the form of Eq.~(\ref{ffun}) is a periodic modulation in one direction, i.e. the stripe solution. 
When allowing fluctuations in the phase of this solution, the local order parameter takes the form 
\begin{equation}
\phi(\vec{x})=c_0+\sum_{n=1}^{\infty}c_n\cos(nk_0x+nk_0u(\vec{x})) 
\end{equation}
where $u(\vec{x})$ represent the deformation field of the stripes and $x$ is the direction transversal to the the average orientation of the stripes. Substituting the proposed anzats in the variational free energy the elastic effective Hamiltonian, up to second order in $\hat{u}(\vec{k})$, is\cite{AbKaPoSa1995,MeStNi2015,ToNe1981,MeStNi2016} 
\begin{equation}
\mathcal{F}[\phi] = \mathcal{F}[\phi_0]+\frac{1}{2}\int \frac{d^2\vec{k}}{(2\pi)^2}\hat{B}(\vec{k})\hat{u}(\vec{k})\hat{u}(-\vec{k}),
\label{elHam}
\end{equation}
where
\begin{equation}
\label{disRel}
 \hat{B}(\vec{k})=B(k_x^2+\lambda^2k_y^4),
\end{equation}
and 
\begin{align}
  \begin{split}
  B &= \frac{1}{2!}\frac{\partial^2}{\partial k_x^2}\left(\sum_{n=1}^{\infty}\frac{c_n^2n^2k_0^2}{4}\left[\hat{A}(\vec{k}-n\vec{k}_0)+\hat{A}(\vec{k}+n\vec{k}_0)\right]\right)\bigg\rvert_{\vec{k}\rightarrow0}\\
B\lambda^2 &= \frac{1}{4!}\frac{\partial^4}{\partial k_y^4}\left(\sum_{n=1}^{\infty}\frac{c_n^2n^2k_0^2}{4}\left[\hat{A}(\vec{k}-n\vec{k}_0)+\hat{A}(\vec{k}+n\vec{k}_0)\right]\right)\bigg\rvert_{\vec{k}\rightarrow0}.
\end{split}
  \label{eqB}
\end{align}

The quadratic effective Hamiltonian of the deformation field given by Eq.~(\ref{elHam}) is an exact and general result, valid for any stripe pattern. The elastic stiffnesses $B$ and $B\lambda^2$ can be numerically determined from the values of $c_n$ and $k_0$ characterizing the stripe pattern minimizing the variational free energy of Eq.~(\ref{ffun}). It is worth noticing that setting the deformation field of the stripes $u(\vec{x})$ as a constant, the effective Hamiltonian Eq.~(\ref{ffun}) remains invariant as expected. 

The elastic Hamiltonian given in Eq.~(\ref{elHam}) is just the starting point of the analysis of the topological melting of a system of stripes as discussed by Toner and Nelson\cite{ToNe1981}. The main consequence of the specific form of the obtained elastic Hamiltonian is the divergence of $\langle u^2(\vec{x})\rangle$, which implies that the stripe order is unstable in two-dimensions. 
At the same time, a careful analysis of the fluctuations of the stripe orientation $\theta(x)=\partial_yu(x)$ allows to conclude that, at low temperature, the system have a phase with quasi-long-range orientational order, characterized by power-law decaying orientational correlations. As shown by Toner and Nelson\cite{ToNe1981}, the establishment of such a weak orientational order at low temperature is a consequence of the presence of topological defects (dislocations) in the stripe structure. This scenario leads to an effective orientational Hamiltonian of the form of a XY model, given by:
\begin{equation}
  \Delta H_{ef}=\frac{1}{2}\int d^2x\ K(\vec{\nabla} \theta(\vec{x}))^2,
  \label{xy1}
\end{equation}
where the effective orientational stiffness ($K$) is determined to be\cite{ToNe1981}
\begin{equation}
 K=\frac{B\lambda^2+2E_d}{2},
\end{equation}
where $E_d$ represents the energy of a dislocation, which in our case can be estimated\cite{KaPo1993} as $E_d=\pi B\lambda^{1/2}k_0^{-3/2}$. The RG equations for the XY effective model, Eq.~(\ref{xy1}), are known to be
\begin{align}
  \begin{split}
 \frac{dK}{dl} &=-4\pi^3K(l)^2y^2(l)\\
 \frac{dy}{dl} &=(2-\pi K(l))y(l)
 \end{split}
 \label{xyRen}
\end{align}
where the bare quantities are given by $y(0)=\exp(-E_c/k_BT)$ and $K(0)=K/k_BT$. 
The energy of the relevant topological defects (disclinations) can be estimated by a technique equivalent to that used by Kashuba and Pokrovsky\cite{KaPo1993} to be $E_c=K\pi(\gamma-\mathrm{cosintegral}(\pi)+\ln(\pi))$.

The calculation of the renormalized stiffness $K(+\infty)$ is possible by solving the set of ordinary differential equations of Eq.~(\ref{xyRen}) with proper initial conditions, that ultimately can be written in terms of the bare stiffnesses $B$ and $\lambda$ given in Eq.~(\ref{eqB}). As it is well established in the KTHNY theory, the disordered phase is identified as the set of points at which $K(+\infty)$ is zero. 
This allows the identification of the transition between a quasi-long-range orientationally ordered stripe phase and the stripe liquid, characterizing what is called as {\it nematic transition}\cite{BaSt2007}.


\subsection{Topological characterization of the bubbles phase}

Now the equivalent analysis is described for the bubble configurations.      
An analytical study of the topological order of the bubble pattern, to the best of our knowledge, is not currently available in the literature. 
Nevertheless, the same principles of the classical melting theory of stripes can be generalized to this case. 

We start by deducing the elastic Hamiltonian of the bubble pattern in the limit of small deformations. 
In this case we admit that the deformation field must be a two component vectorial field.
In order to allow only deformation fields that respect locally the symmetries of the bubble pattern we consider the following form of the modulation in the presence of phase fluctuations:
\begin{equation}
\phi(\vec{x})=c_0+\sum_{n=1}^{\infty}c_n\cos(k_0\vec{b}_n\cdot\vec{x}+k_0\vec{b}_n\cdot\vec{u}(\vec{x})), 
\end{equation}
where $\vec{u}(\vec{x})$ represents the deformation field of the bubble configuration.
Following a procedure similar to the one described in the previous subsection it is straightforward to reach the following elastic Hamiltonian for the bubble system:
\begin{align}
\nonumber
\mathcal{F}[\phi]=&\mathcal{F}[\phi_0]+\frac{1}{2}\int \frac{d^2\vec{k}}{(2\pi)^2}((2m+l)k_x^2+mk_y^2)\hat{u}_x(\vec{k})\hat{u}_x(-\vec{k})\\
&+\frac{1}{2}\int \frac{d^2\vec{k}}{(2\pi)^2}((2m+l)k_y^2+mk_x^2)\hat{u}_y(\vec{k})\hat{u}_y(-\vec{k})
+\frac{1}{2}\int \frac{d^2\vec{k}}{(2\pi)^2}(2m+2l)k_xk_y\hat{u}_x(\vec{k})\hat{u}_y(-\vec{k}),
\label{Hefbub}
\end{align}
where the elastic stiffnesses $m$ and $l$ can be calculated by solving the following system of equations:
\begin{align}
\begin{split}
 2m+l&=\frac{1}{2!}\frac{\partial^2}{\partial k_x^2}\left(\sum_{n=1}^{\infty}\frac{c_n^2k_0^2(b_n^x)^2}{4}\left[\hat{A}(\vec{k}-k_0\vec{b}_n)+\hat{A}(\vec{k}+k_0\vec{b}_n)\right]\right)\bigg\rvert_{\vec{k}\rightarrow0},\\
 m&=\frac{1}{2!}\frac{\partial^2}{\partial k_y^2}\left(\sum_{n=1}^{\infty}\frac{c_n^2k_0^2(b_n^x)^2}{4}\left[\hat{A}(\vec{k}-k_0\vec{b}_n)+\hat{A}(\vec{k}+k_0\vec{b}_n)\right]\right)\bigg\rvert_{\vec{k}\rightarrow0}.
 \end{split}
 \label{Lame}
\end{align}

The effective Hamiltonian given by Eq.~(\ref{Hefbub}) has the form of the elastic Hamiltonian of a triangular crystal in two dimensions and as expected the free energy of the pattern remains invariant if the field $\vec{u}(\vec{x})$ is set to a constant. 
This implies that the triangular array of bubbles can be treated like a triangular lattice of particles, with Lam\'e's elastic coefficients given by the expressions in Eq.~(\ref{Lame}). Consequently, the topological properties of the bubbles lattice can be studied using the results from KTHNY theory for the two-dimensional melting of triangular lattice of particles.  

This theory predicts that the triangular lattice of particles melts by a sequence of two transitions as we increase temperature. 
At low temperatures, the triangular lattice of particles (bubbles) find itself in a state where the positional correlation decays as a power-law and the next-neighbor bond angle presents long-range order, characterizing the 2D solid phase. 
Upon increasing the temperature a transition is predicted to an intermediate phase usually called the hexatic phase, characterized by exponentially decaying positional correlations and power-law bond angle orientational correlations. 
Finally a further increase of the temperature leads the system to a liquid state in which the positional and the bond angle orientational order are both short ranged.     

\subsubsection{Solid-hexatic transition}

The transition between the 2D solid and the hexatic phases is mediated by the proliferation of dislocation pairs\cite{Ko2016}. The effective Hamiltonian of the interacting dislocation field is given by the expression:   
\begin{equation}
 \Delta H_D=-\frac{K_d}{8\pi a^4}\int d^2x\ d^2x'\left[\vec{b}(\vec{x})\cdot\vec{b}(\vec{x}')\ln\left(\frac{\vert\vec{x}-\vec{x}'\vert}{a}\right)-
 \frac{\vec{b}(\vec{x})\cdot(\vec{x}-\vec{x}')\vec{b}(\vec{x}')\cdot(\vec{x}-\vec{x}')}{\vert\vec{x}-\vec{x}'\vert^2}\right]+E_D\int\ \frac{d^2x}{a^2}\vert\vec{b}(\vec{x})\vert^2,
 \label{Edis}
\end{equation}
where $a$ represents the short distance cut-off of the interacting dislocation problem, which could be interpreted as 
the minimum possible distance in a dislocation pair. 
The elastic constant $K_d$ is given by $\frac{4m(m+l)}{(2m+l)}\left(\frac{4\pi}{\sqrt{3}k_0}\right)^2$ and the parameter $E_D$ represents the energy of a dislocation. 
Since the energy of an isolated dislocation diverges in the 2D solid phase, $E_D$ must be understood as half the minimum energy of a dislocation conjugate pair. 
To estimate this value we apply a technique similar to the one used by Kashuba and Pokrovsky \cite{KaPo1993}. 
First, the interacting part of the Hamiltonian of Eq.~(\ref{Edis}) is written in Fourier space, and a configuration of Burguer's vectors of the dislocation pair is selected by minimizing the interaction energy.
Then the calculation is performed directly in Fourier space properly selecting the momentum cut-off. 
This procedure leads to the result $E_D=K_d(\gamma-\mathrm{cosintegral}(\pi)+\ln(\pi))/(8\pi)$. We can then use the 
RG equations for the renormalized stiffness in the 2D solid phase within the two-dimensional KTHNY theory, given by:   
\begin{align}
 \frac{dK_s(l)}{dl}&=-K_s^2(l)\left[\frac{3\pi}{2}y(l)^2\exp(K_s(l)/8\pi)I_0(K_s(l)/8\pi)-\frac{3\pi}{4}y(l)^2\exp(K_s(l)/8\pi)I_1(K_s(l)/8\pi)\right]\\
 \frac{dy(l)}{dl}&=\left(2-\frac{K_s(l)}{8\pi}\right)y(l)+2\pi y(l)^2\exp(K_s(l)/8\pi)I_0(K_s(l)/16\pi),
\end{align}
where
$K_s(0)=K_d/k_BT$ and $y(0)=\exp(-E_D/k_BT)$. 
The integration of this system of differential equations allows to identify when the transition from the 2D solid to the hexatic phase takes place. 
We define the 2D solid phase region as composed by those points satisfying that $K_d(+\infty)$ is finite. 
The integration of the RG equations in the regions where the positional order is only local ($K_d(+\infty)=0$) allows the determination of the positional correlation length in units of the short distance cut-off ($\xi_p/a$). 
As it is well established in the RG theory $\xi_p/a$ can be calculated as $\exp(l^*)$, where $l^*$ represents the RG time to $K_d(l)$ reach $K_d(+\infty)$. 
Normally $K_d(l)$ have a very steep evolution so it is relatively easy to identify $l^*$. 
In this way the positional correlation length can be calculated, which as discussed below is a central quantity for the characterization of the orientational order.   

\subsubsection{Hexatic-liquid transition}
As we mentioned before, once the 2D solid phase melts, the system can develop an hexatic or a liquid phase. 
In this region of the phase diagram the orientational order, according to the classical theory of melting, can be effectively described by a Hamiltonian of the form:
\begin{equation}
 \Delta H_{o}=\frac{1}{2}\int d^2x\ K_A(\vec{\nabla} \theta(\vec{x}))^2,
 \label{XYo}
\end{equation}
where the so-called Frank's constant $K_A$ is normally estimated\cite{NeHa1979,Ko2016} to be $2E_D(\xi_p/a)^2$. 
Since the effective model describing the orientational order in the long wave limit coincides with the $XY$ model, as can be seen in Eq.~(\ref{XYo}), we can take advantage of the well known RG equations for this model to evaluate the renormalized Frank's constant ($K_h(l)$). These equations have the form:
\begin{align}
\begin{split}
 \frac{dK_h(l)}{dl}&=-4\pi^3K_h(l)^2y^2(l)\\
 \frac{dy(l)}{dl}&=(2-\pi K_h(l))y(l),
 \end{split}
\end{align}
where the bare elastic constant $K_h(0)$ is given by $K_A/k_BT$ and the bare fugacity $y(0)$ is taken as usual $\exp(-E_A/k_BT)$, with $E_A$ representing the energy of the relevant topological defects (disclinations) of this transition. 
The energy of the disclinations can be estimated in a form equivalent to the one used for calculating the energy of a dislocation in the previous section. Following that procedure we reach to $E_A=K_A\pi(\gamma-\mathrm{cosintegral}(\pi)+\ln(\pi))/36$. 
This result completes the set of equations for characterizing the orientational order of the bubbles phase. 
The set of equations obtained in this section allows to identify the melting temperature of the hexatic phase as the one in which the renormalized orientational stiffness $K_h(+\infty)$ first goes to zero.

\section{Topological phase diagrams}

The numerical procedure for the construction of the topological phase diagram in the $H$-$T$ plane can be outlined as follows. 
First the free energy functional of Eq.~(\ref{ffun}) is minimized, which allows to find the optimal stripes and bubbles solution. 
These results are then used to calculate the Lam\'e's coefficients ($B$, $B\lambda^2$, $m$, $2m+l$) for the stripes and bubbles. 
The knowledge of these quantities is the starting point for the application of the two-dimensional classical theory of melting described in the previous sections, which ultimately allows to establish the topological order for each local pattern. 

Following this procedure we construct several $H$-$T$ phase diagrams varying the curvature of the fluctuation spectrum $a$ in Eq.~(\ref{ffun}). 
As can be noticed in the definition of the model, different values of the parameter $a$ are equivalent to different relative strengths of the dipolar interaction $\delta^{-1}$ in the microscopic Hamiltonian of Eq.~(\ref{Hmic}). 
As studied in previous works, varying the curvature of the fluctuation spectrum this system develops a strong reentrance of the homogeneous phase with temperature\cite{MeNiDi2019}. 
However, to the best of our knowledge this type of topological phase diagrams have not been obtained before, and consequently a number of interesting questions like the influence of the reentrance in the topological properties of the system remains open.

\begin{figure}[th!]
\begin{center}
\includegraphics[width=0.39\columnwidth]{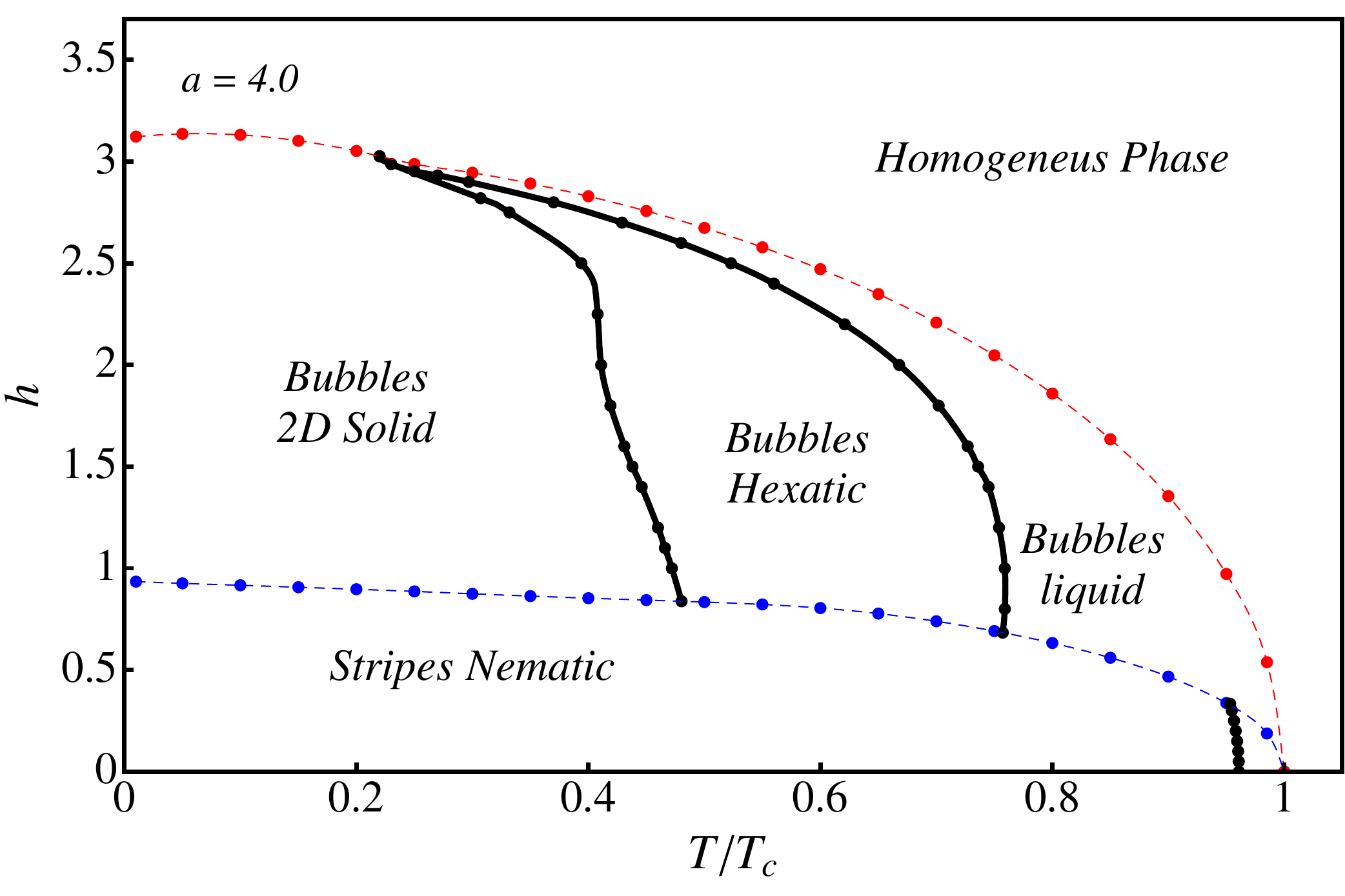}
\includegraphics[width=0.39\columnwidth]{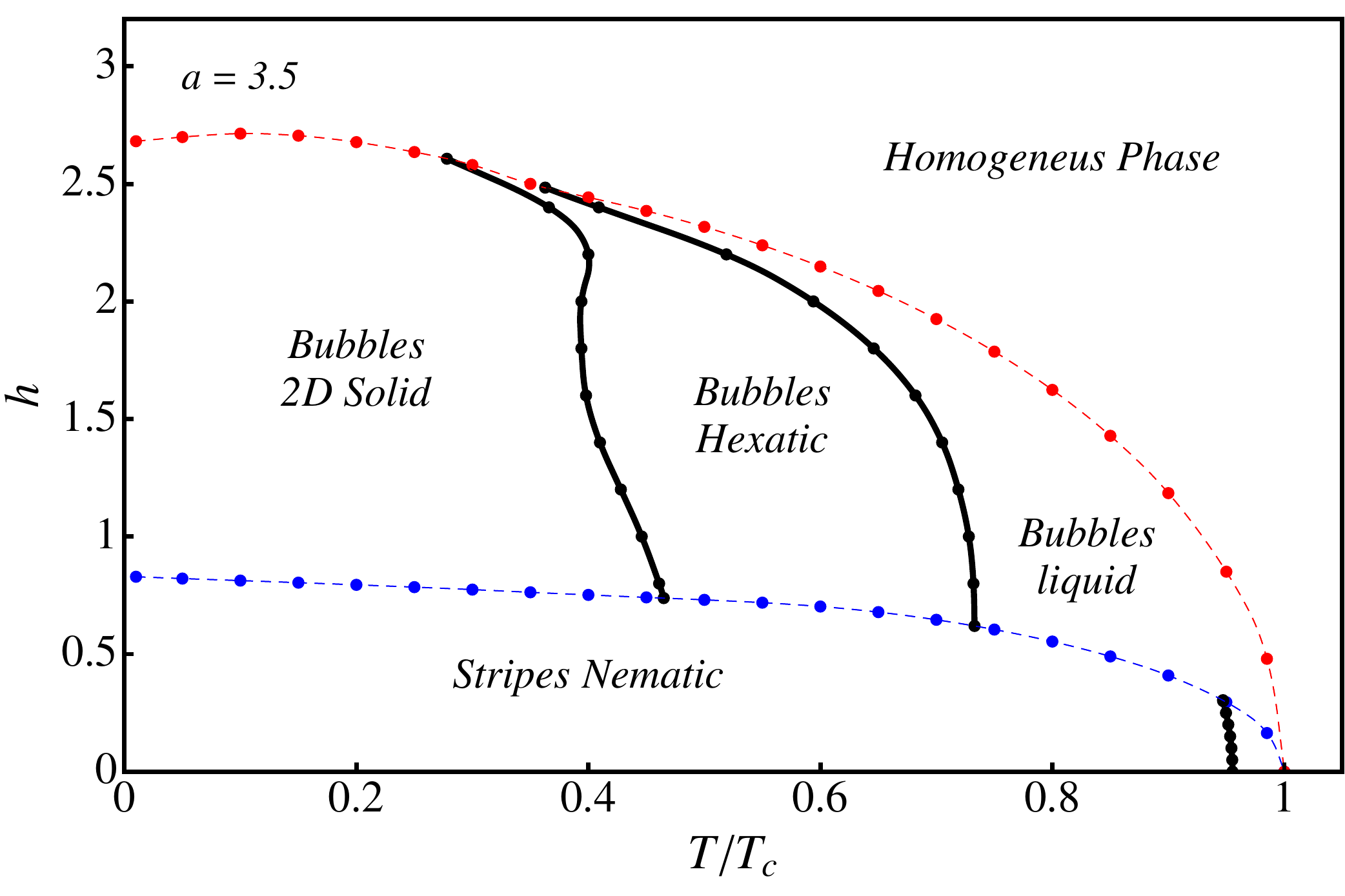}\\
\includegraphics[width=0.39\columnwidth]{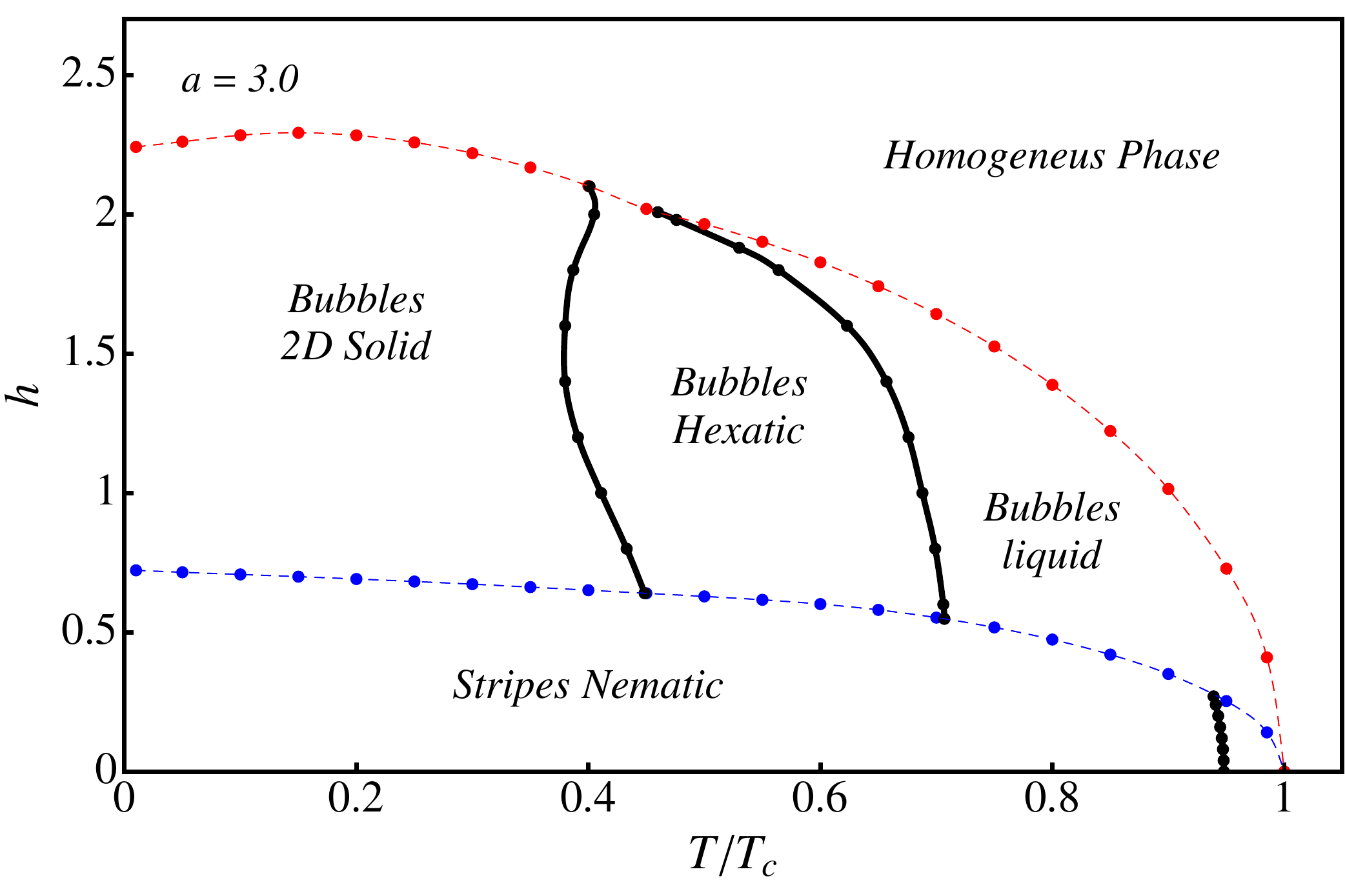}
\includegraphics[width=0.39\columnwidth]{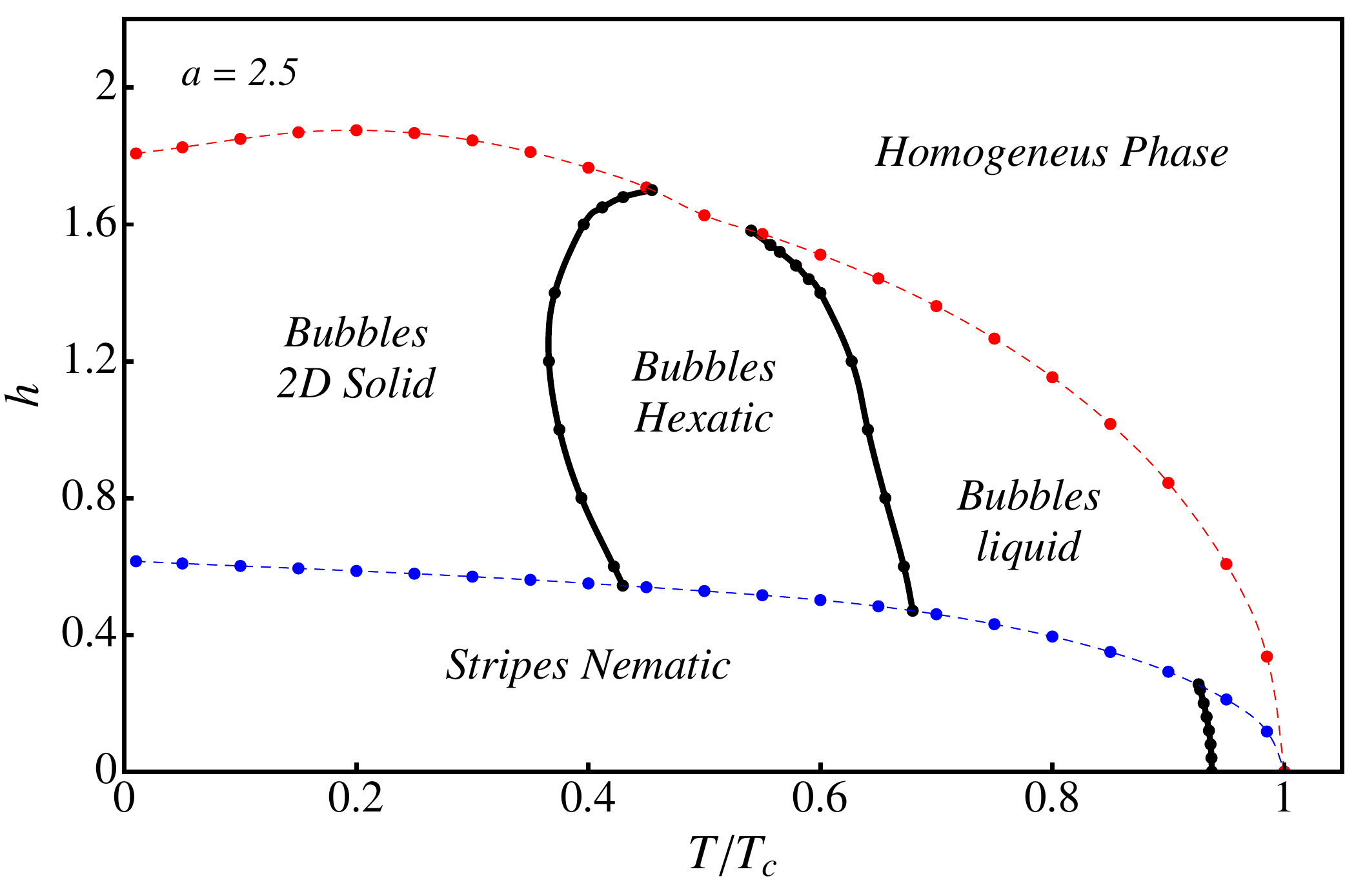}\\
\includegraphics[width=0.39\columnwidth]{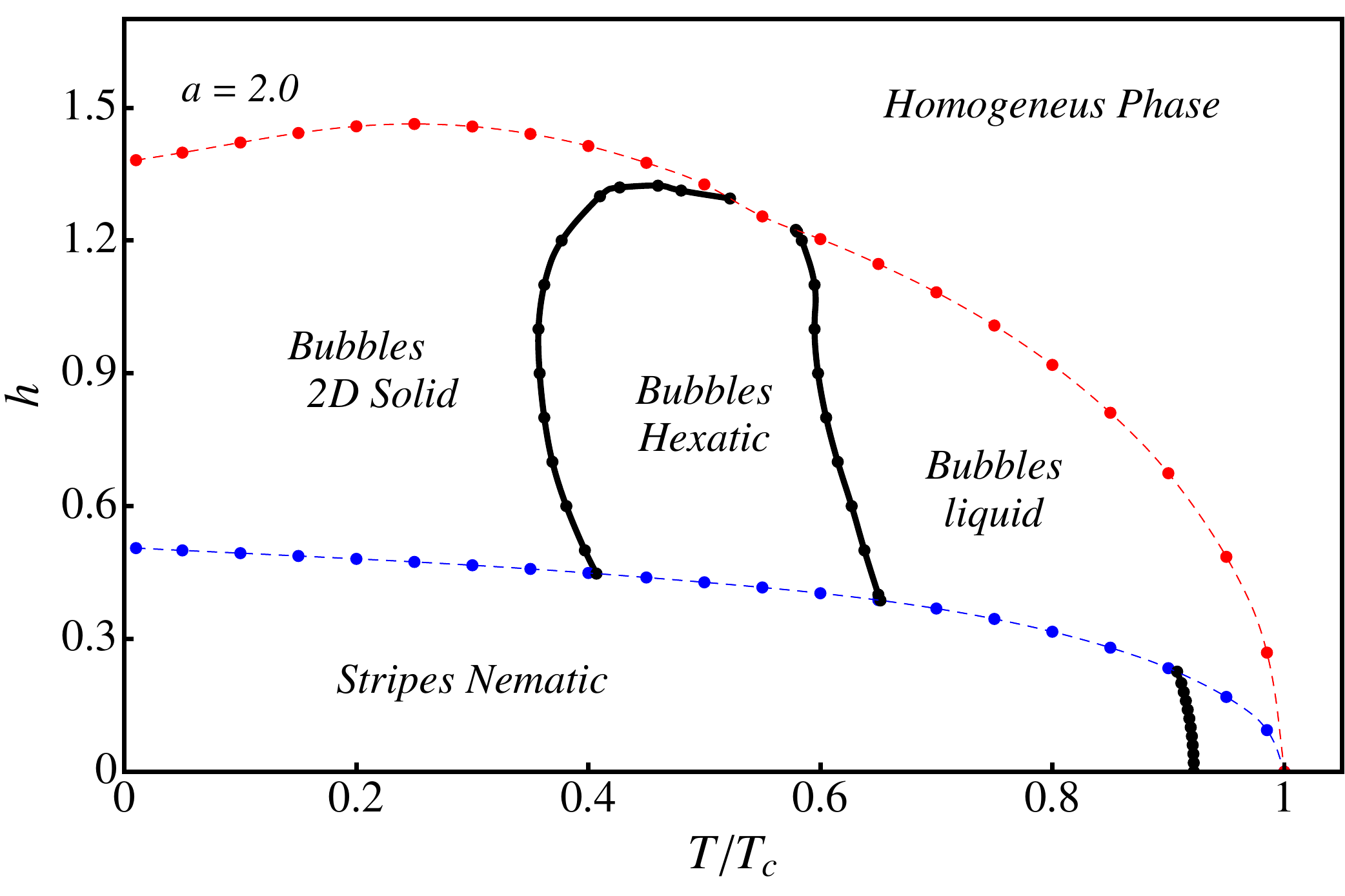}
\includegraphics[width=0.39\columnwidth]{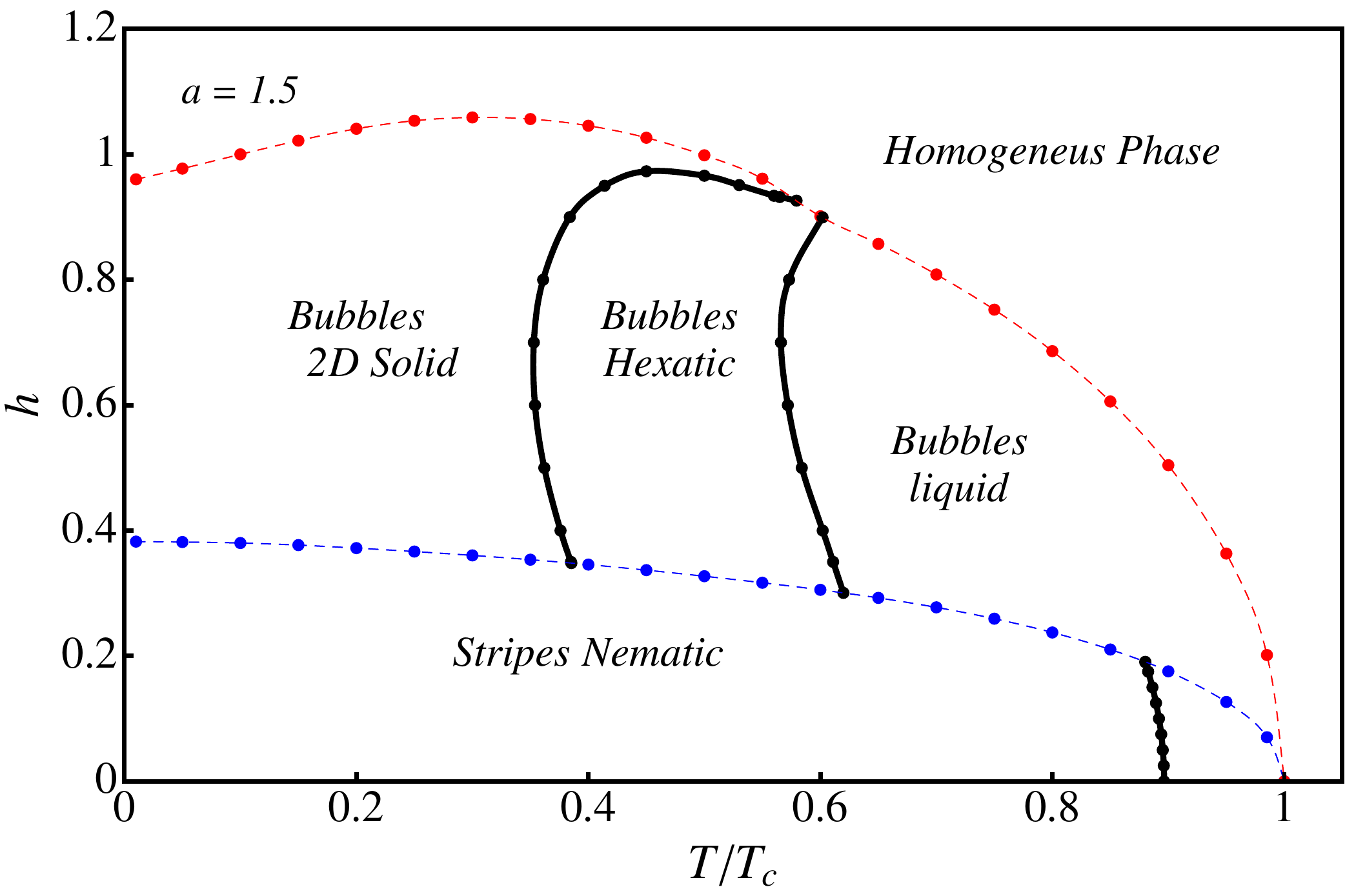}\\
\includegraphics[width=0.39\columnwidth]{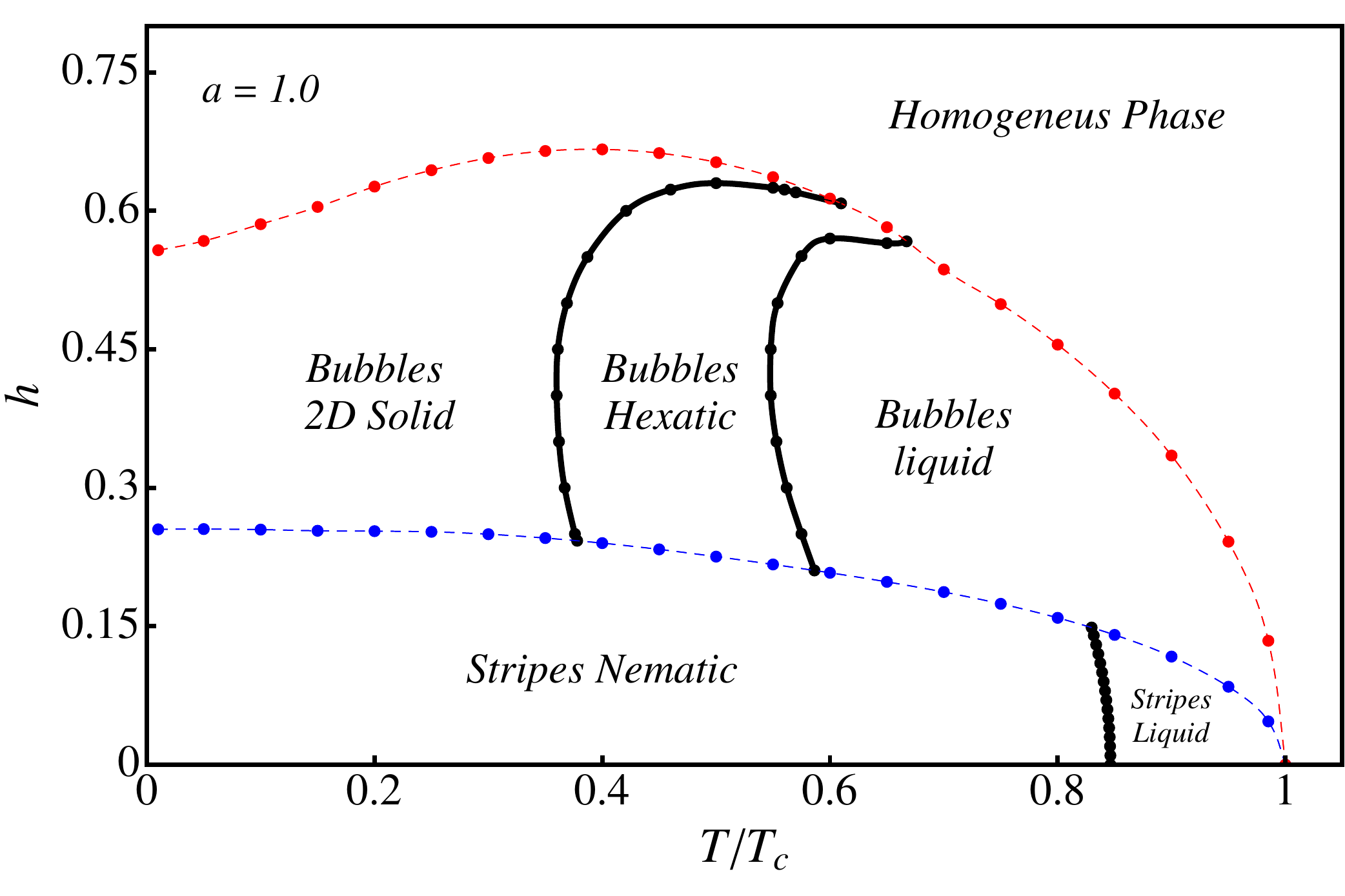}
\includegraphics[width=0.39\columnwidth]{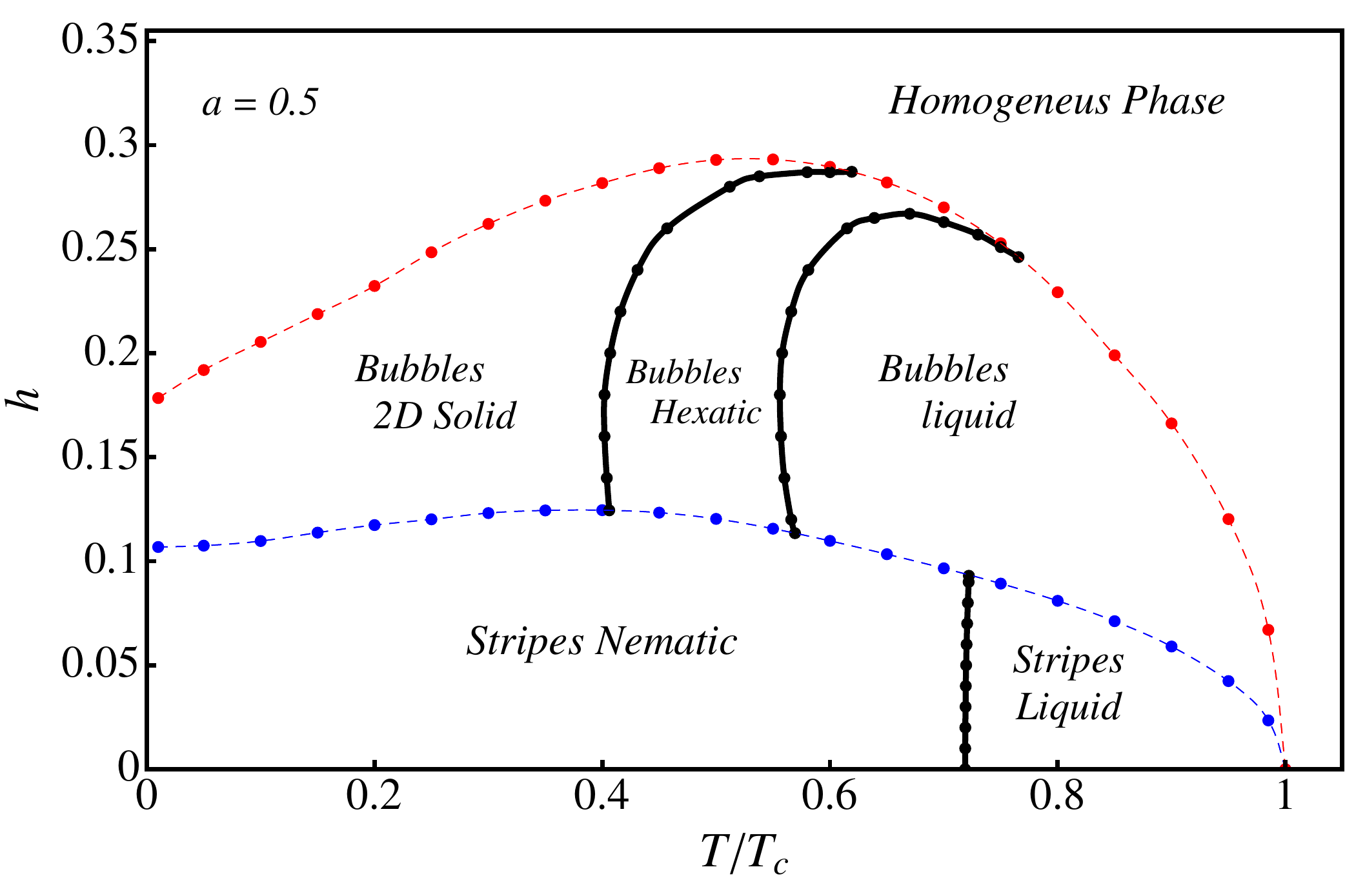}\\
\end{center}
\hspace{-4.3cm}
\caption{
$H$-$T$ topological phase diagrams varying the curvature of the fluctuation spectrum $a$. 
From top to bottom and left to right the values are $a={4.0,3.5,3.0,2.5,2.0,1.5,1.0,0.5}$. The points represent the numerical results in all cases, and the lines are guides to the eye. Full black curves identify lines of critical points, while the dashed red and blue curves, resulting from mean-field calculations, identify first-order transitions.
In all phase diagrams three phases with broken symmetries can be identified: nematic,  2D solid and hexatic. All other phases are topologically equivalent since the rotational and translational symmetries are preserved.
As discussed in the text, we expect that the transitions between liquid phases and the homogeneous phase are not actual phase transitions, just crossovers.
}
\label{fig1}
\end{figure}

In Fig.~\ref{fig1} a sequence of $H$-$T$ topological phase diagrams is presented varying the curvature of the fluctuation spectrum $a$ from a regime with practically zero reentrance to a regime of strong reentrance. 
The first thing we notice for all diagrams is that, depending on the value of the magnetic field, the sequence of transitions is not always the one predicted by the classical theory of melting. 
This happens not only due to the relative change of stability between the modulated phases, but also because of the occurrence of direct transitions from the modulated to the disordered phase (homogeneous phase). 
This kind of transitions are beyond the scope of the classical theory of two-dimensional melting and are obtained here as a result of the simultaneous use of a density functional theory and the RG techniques, providing a richer outcome than that produced by each technique separately.     

In all phase diagrams we distinguish six different regions. Three are phases with broken symmetries, namely the {\it stripes nematic}, the {\it bubbles 2D solid} and the {\it bubbles hexatic} phases. The other three regions ({\it stripes liquid}, {\it bubbles liquid} and {\it homogeneous phase}) are equivalent in terms of their symmetries, and consequently there is no symmetry breaking between these regions. Based also on previous results from numerical simulations \cite{NiSt2007, NiMeSt2016, MeNiDi2019}, we expect that the transitions between these regions are not actual phase transitions, just crossovers.

The sequence of phase diagrams in Fig.~\ref{fig1} shows that decreasing the relative strength of the dipolar interaction $\delta^{-1}$, the development of reentrance in the phase diagram produces continuously a qualitative transformation of the topological phases, specifically of the boundary between different phases. 
In the absence of reentrance, for high values of $a$, the boundaries between the topological phases are given by monotonous decreasing curves in the $H$-$T$ plane. 
This is the behavior expected in a normal phase diagram since the intuition suggests that the application of an external magnetic field weakens the modulations and consequently these phases are less pronounced. 
However, we observe that, as the homogeneous reentrance becomes stronger, the boundaries of the topological phases develop a non-monotonic behavior.
This tendency is responsible for the eventual appearance of unexpected reentrances of the topological phases when the magnetic field is increased at constant temperature. 
See for instance the phase diagram presented for the lowest value of the parameter $a$ ($a=0.5$), in this case increasing the magnetic field a reentrance appears in the boundaries of all topological phases.          

In our level of approximation, all topological transitions within the bubbles or stripes mean-field regions occur at a fixed value of the relevant elastic constants normalized by the thermal energy, $K(0)$, $K_s(0)$ and $K_h(0)$. 
If we trace back the dependence of these quantities with the microscopic parameters we find a relation of proportionality with the microscopic elastic coefficients ($B$, $B\lambda^2$, $m$, $2m+l$) and with the modulation length ($2\pi/k_0$). 
A numerical analysis of the behavior of these quantities in the region of the boundaries of the topological phases reveals that in all cases, as expected, when the external field is increased, the microscopic elastic coefficients decrease as modulations become less pronounced. 
At the same time, however, the modulation length of the patterns is a quantity that increases as we raise the external field. 
This scenario suggests that, in the presence of a reentrance of the homogeneous phase, the non trivial behavior of the topological phase boundaries is mainly a consequence of the behavior of the modulation length when increasing the external field. 
We identify that for these topological reentrances the modulation length of the pattern growths with the external field fast enough to overcome the effect of the decreasing microscopic elastic coefficients, making finally the relevant normalized elastic constants ($K(0)$, $K_s(0)$, $K_h(0)$) to increase with the external field.      

At this point, is worth to discuss which are our expectations for the topological phase diagrams of the system beyond our perturbative treatment. As can be appreciated in Fig.\ref{fig1} the boundaries of the topological phases determined by the combined application of the classical theory of melting and the mean field techniques produces cusps at the intersection of the phase boundaries determined through each different method. More than this, in the case of low values of $a$, very sharp ends of the hexatic and 2D solid phases results also as an effect of the calculation method employed. We understand that although the microscopic mechanism producing the topological reentrances is robust, as explained above, the discontinuities in the derivative of the $H$vs.$T$ phase boundaries are certainly an artifact of the perturbative method. In contrast, we expect that the actual phase diagrams still present the described topological reentrances but with smooth and probably more round phase boundaries, as well as lower critical temperatures considering the underlying mean-field and pertubative treatment of the fluctuations.

\section{Critical properties of the topological transitions}

Although the critical properties are well understood in the context of KTHNY theory, considering temperature as the tuning parameter, no predictions are available for the phase transitions varying parameters like the magnetic field $H$ and the fluctuation spectrum curvature $a$. It is worth noticing that varying these parameters do not break explicitly any symmetry of the system. 

\begin{figure}[th!]
\begin{center}
\includegraphics[width=0.52\columnwidth]{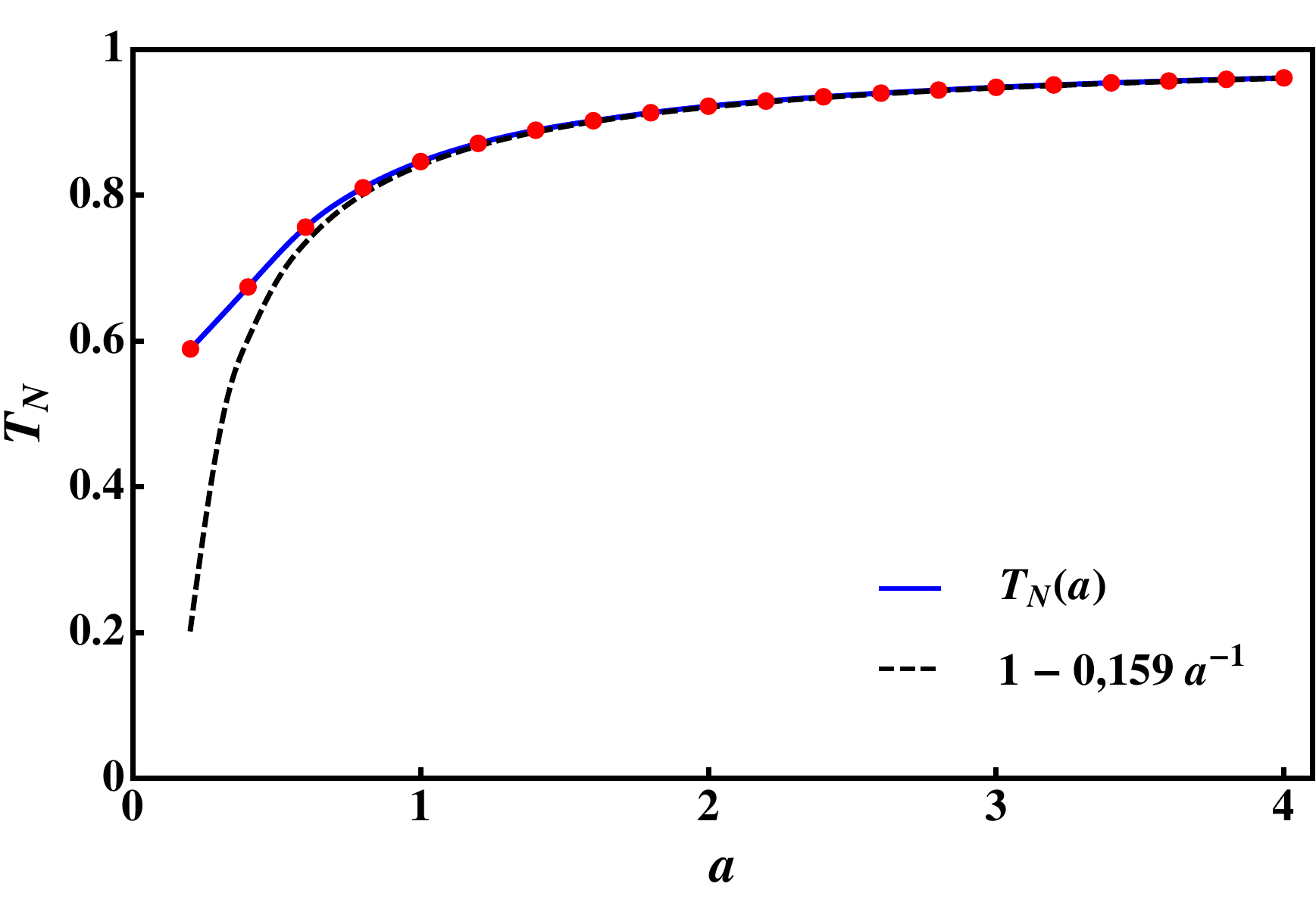}
\caption{Nematic critical temperature $T_N$ at zero external field as a function of the curvature of the fluctuation spectrum $a$.
  Dots represent the numerical results, while the dashed line corresponds to the best fit by a functionality of the form $T_N(a)=1-ca^{-1}$.
	For large values $a\to\infty$ the nematic transition approaches the mean-field critical temperature. 
}
\label{fig2}
\end{center}
\end{figure}

A first inquiry that one may ask would be about the dependence with $a$ of the value of the critical nematic temperature at zero magnetic field $T_N$, which is shown in Fig.~\ref{fig2}. The obtained results confirms the expectation that increasing the curvature of the fluctuation spectrum, maintaining the position of the minimum $\hat{A}(k)$, increases the value of $T_N$. In the limit of large values of $a$ the nematic transition temperature approaches the mean field critical temperature.

Considering that in our effective model exists only one free parameter ($a$), the curve presented is universal, since any microscopic Hamiltonian like that of  Eq.~(\ref{Hmic}) can be mapped into an effective Hamiltonian of the form of Eq.~(\ref{ffun}) and will correspond to a single value of $a$ and $\hat{A}(k_0)$.

Regarding the characterization of the form of the nematic phase boundary (nematic to stripe liquid) at low external fields, we found that the critical field $h_c$ varies with temperature following a law of the form $h_c\propto\vert T_c-T\vert^{1/2}$. 
This kind of behavior can be understood considering that, as discussed before, the topological transitions occur at specific values of the relevant stiffness normalized by the thermal energies ($K(0)$, $K_s(0)$, $K_h(0)$). 
In the case of the nematic transition, the fact that this transition occurs within the stripe region allows to conclude that $K(0,H,T)$ have to be an even function of $H$, so that  $K(0,H,T)=K(0,-H,T)$. 
This property can be deduced once it is recognized that the mean-field free energy is an even function of $H$. 
In the region of low fields we can expand $K(0,H,T)$ around the point $(T_N,0)$ obtaining
\begin{equation}
K(0,H,T)=K(0,0,T_N)+\partial_TK(0,0,T_N)(T-T_N)+\frac{1}{2}\partial^2_HK(0,0,T_N)H^2
\end{equation}  
In this way the solution of the equation $K(0,H,T)=K(0,0,T_N)$ will be
\begin{equation}
 H=\left(-\frac{2\partial_TK(0,0,T_N)(T-T_N)}{\partial^2_HK(0,0,T_N)}\right)^{1/2}
 \label{nemHT}
\end{equation}
which explains the form of the observed curves of $H$ versus $T$ for the nematic boundary at low external fields.
 
\begin{figure}[th!]
\begin{center}
\includegraphics[width=0.52\columnwidth]{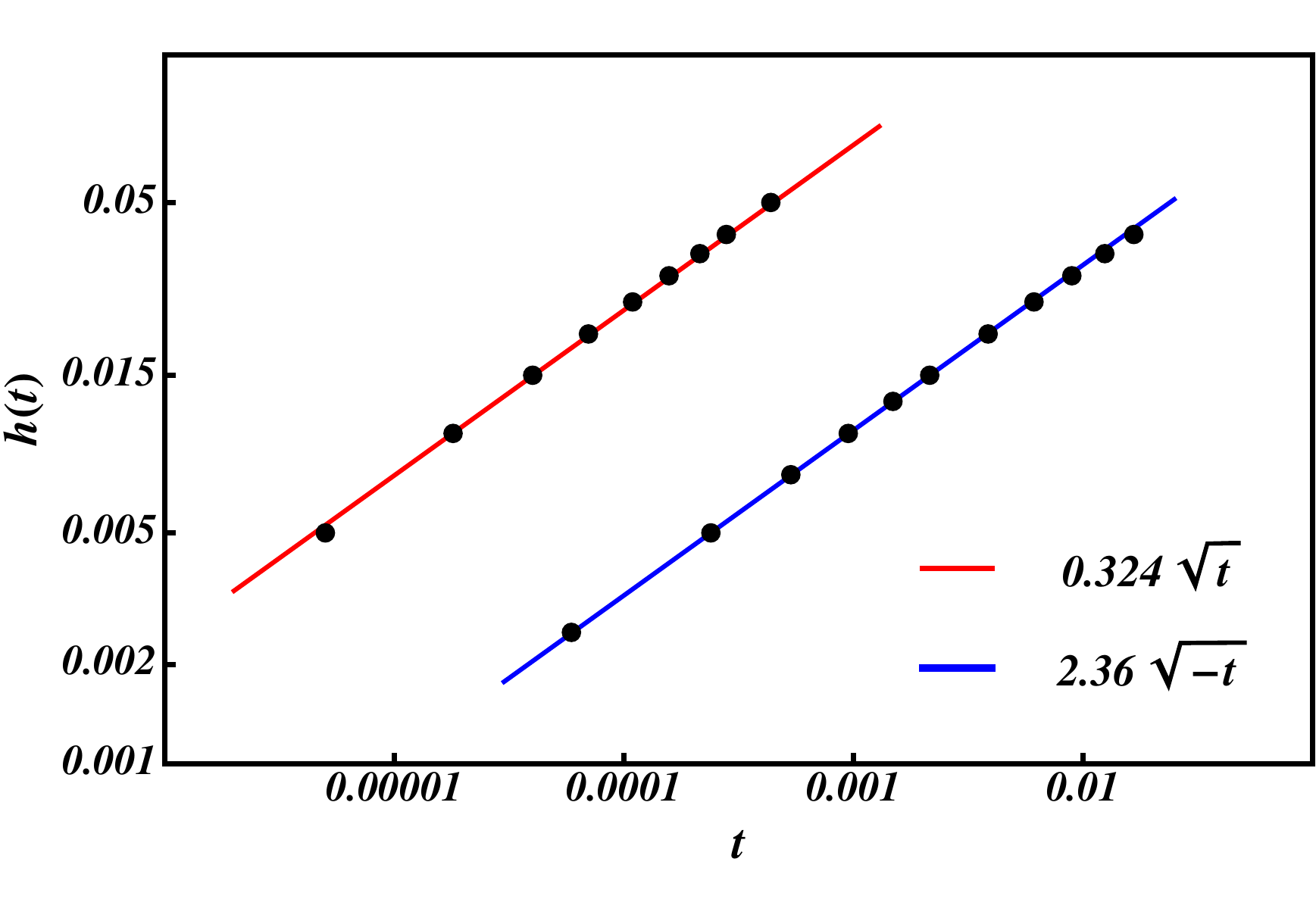}
\caption{
Critical nematic field $h_c(t)$ varying the reduced temperature $\left(t=T-T_N(a)\right)$ for fluctuation spectrum curvatures $a=2.5$ (blue) $a=0.2$ (red). 
For $a=2.5$ ($T_N=0.937529$) there is a normal $h_c(t)$ behavior while for $a=0.2$ ($T_N=0.589172$), reentrance exists varying the external field.    
}
\label{fig3}
\end{center}
\end{figure}

With varying the curvature of the fluctuation spectrum $a$ we observe a change in the shape of the nematic phase boundary (see Fig.~\ref{fig1}). 
In Fig.~\ref{fig3} the nematic boundary is shown for two different cases: one in which the boundary behaves as expected with $h_c(T)$ monotonically decreasing (blue curve), and one with $h_c(T)$ monotonically increasing (red curve), in which a reentrance of the disordered phases appears with varying the external field. 
We realize that even in the anomalous case the field-temperature scaling does not change, as expected from Eq.~(\ref{nemHT}).

An important feature of the boundary between topological phases in the phase diagrams is that they are formed by actual critical points. 
The external field do not break the symmetries of the modulated patterns and consequently we have the possibility of a spontaneous symmetry breaking varying both temperature and magnetic field. 
The critical properties varying temperature are those from KTHNY theory at any point of the boundary between the topological phases. 
However, the critical properties varying the magnetic field are not obvious.    

Now we will show that the critical properties varying the external field is defined by the local form of the topological phase boundary. In what follows, the arguments raised for the nematic phase are valid for the 2D solid and the hexatic phases as well.
Let us consider an arbitrary point of the nematic phase boundary $(H_0,T_0)$. 
Since $K(0,H,T)$ is an analytic function in the $H$-$T$ plane, the series expansion in powers of $(T-T_0)$ and $(H-H_0)$ have the form:
\begin{equation}
 K(0,H,T)=K(0,0,T_N)+\partial_TK(0,H_0,T_0)(T-T_0)+\frac{1}{n!}\partial^n_HK(0,H_0,T_0)(H-H_0)^n,
\end{equation}
where only the leading order terms in $(T-T_0)$ and $(H-H_0)$ are kept and $n$ represents the order of the lowest external field contribution. The relation $K(0,H_0,T_0)=K(0,0,T_N)$ have been used. 
Around the point $(H_0,T_0)$ the solution of the equation $K(0,H,T)=K(0,H_0,T_0)$ have the form:
\begin{equation}
  H-H_0=\left(-\frac{n!\partial_TK(0,H_0,T_0)(T-T_0)}{\partial^n_HK(0,H_0,T_0)}\right)^{1/n}.
 \label{nemHT2}
\end{equation}

At the same time, studying  the variations of $K(0,H,T)$ around $(H_0,T_0)$ by fixing $T$ or $H$, it can be seen that the critical exponents varying the external field $\nu_h$ are related to those varying temperature $\nu_T$ in the form $\nu_h=n\nu_T$. 
This means that the critical properties varying temperature and varying the external field  can be related by just looking at the local form of the topological phase boundary, for any black curve shown in Fig.~\ref{fig1}.

\begin{figure}[th!]
\begin{center}
\includegraphics[width=0.44\columnwidth]{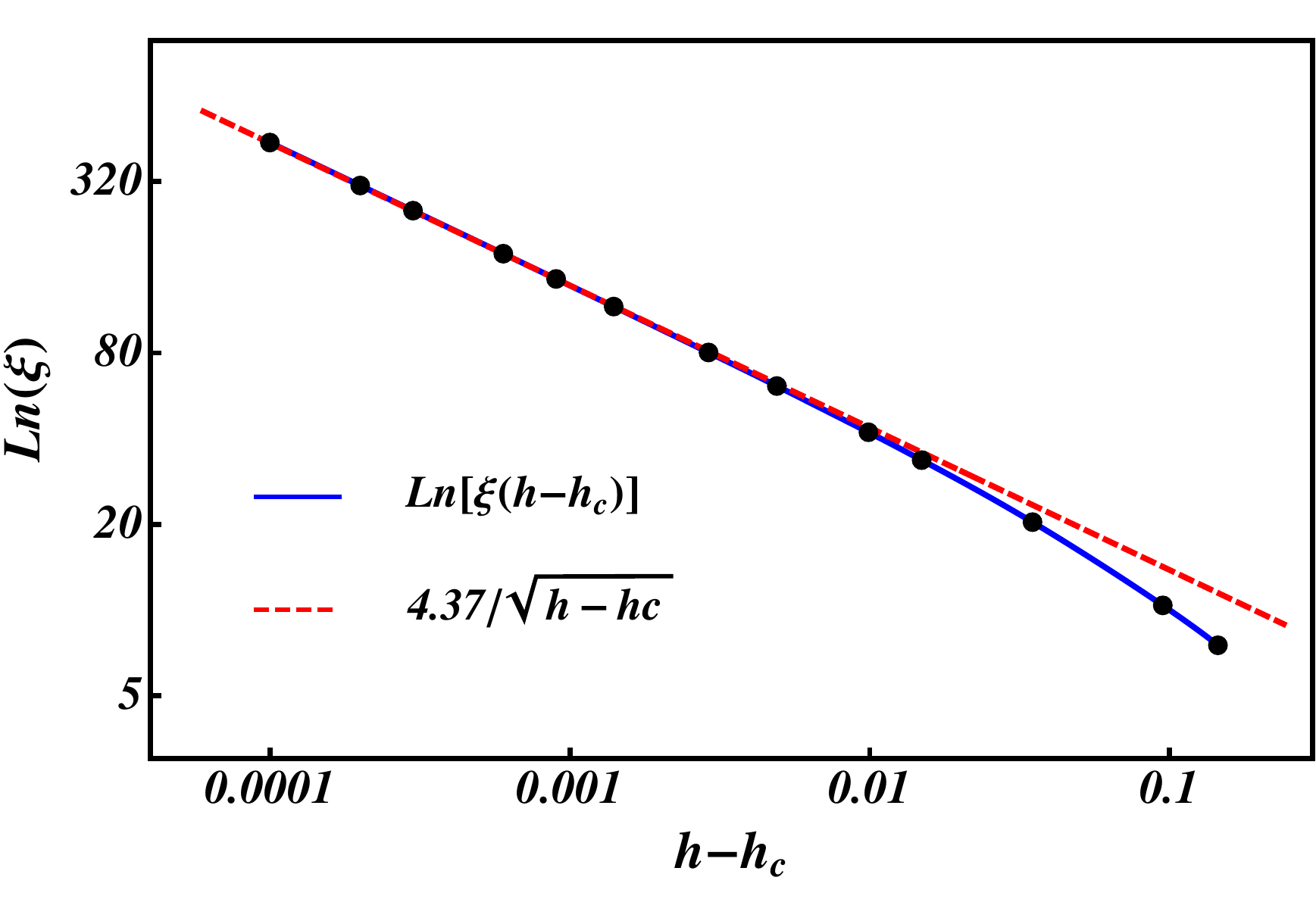}\hspace{0.5cm} 
\includegraphics[width=0.44\columnwidth]{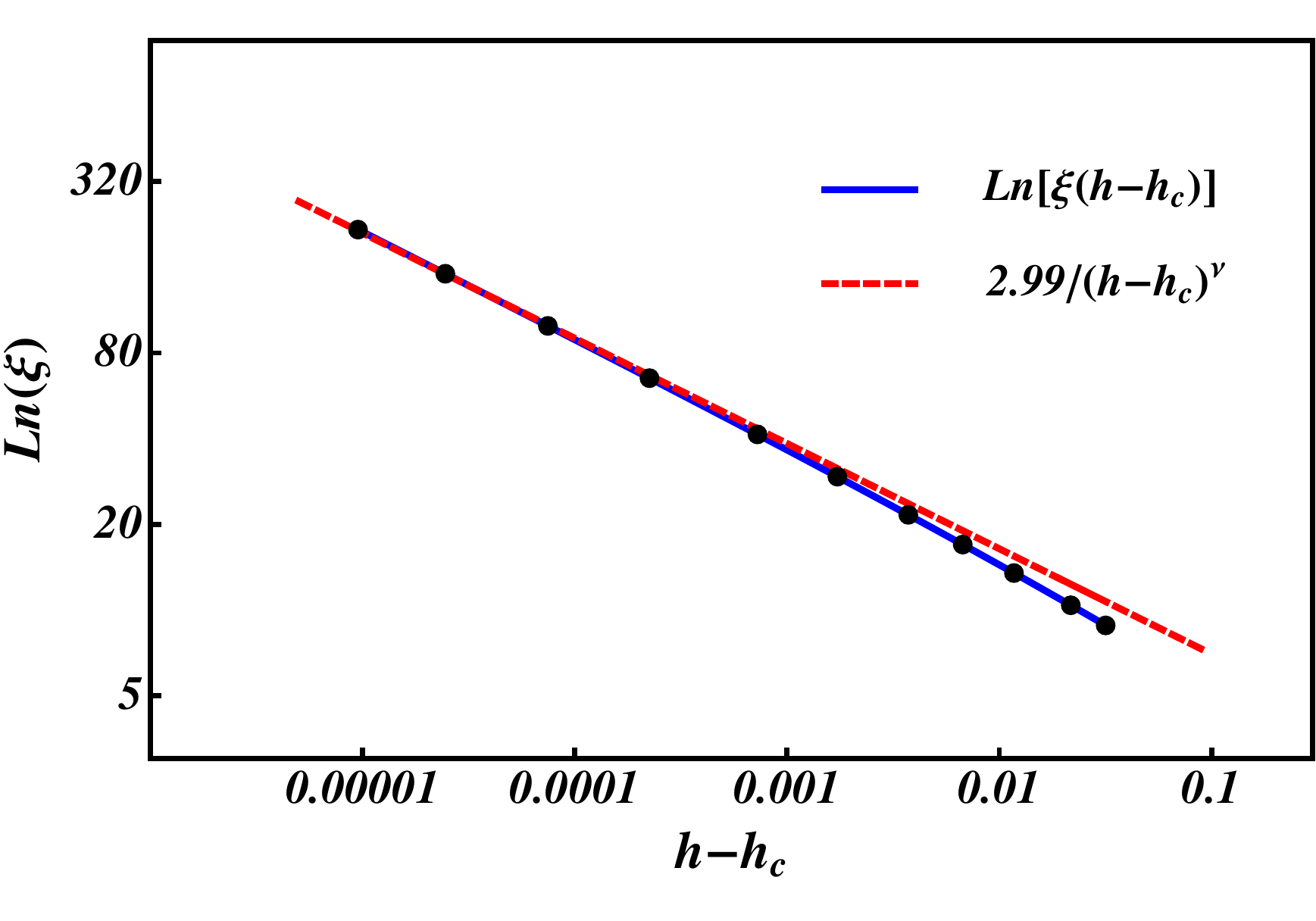}
\caption{ Natural logarithm of the nematic (Left) and solid (Right) correlation lengths as a function of the external field for $a=2.5$, at $T=0.937$ and $T=0.4$ respectively. 
	The estimated critical fields at those temperatures are $h_c=0.0551001$ and $h_c=0.7582755$ respectively. The red lines are fits of the logarithm of the correlation length to the form $A(h-h_c)^{-\nu_h}$, where $A$ is the only fitting parameter. The value of $\nu_h$ for the nematic ($\nu_h=\nu_T=1/2$) and for the 2D solid-hexatic ($\nu_h=\nu_T=0.36963477$) transitions are fixed to those predicted by the classical theory of melting~\cite{NeHa1979}.}
\label{fig4}
\end{center}
\end{figure}

To further illustrate this argument, it is presented in Fig.~\ref{fig4} the numerical results for the nematic and the solid correlation lengths as a function of the external field for two points of the respective phase boundaries - one at the nematic-liquid and other at the solid-hexatic phase boundaries.
Since at the selected critical points the phase boundary is not vertical it is natural to expect  that the critical behavior varying the external field coincides with that varying temperature. This means, according to the classical theory of melting, that the correlation length should behave as $\log{\xi(h)}\propto(h-h_c)^{-\nu_h}$, where $h$ represent the external applied field and $\nu_h$ represent the correlation length critical exponent for K.T. like transitions~\cite{KoTh1973,ToNe1981,NeHa1979} taking the external field as the controle variable. As mentioned before, considering the selected critical points around which the correlation lengths are analyzed, it is expected to have in both cases $\nu_h=\nu_T$. This expectation is confirmed in Fig.\ref{fig4} where the best fit of the correlation length curves shows the same scaling of the classical theory of melting.  

\section{Conclusions}

We have studied the full topological phase diagram of a model of ultrathin dipolar ferromagnetic layer in the limit of strong perpendicular anisotropy. 
These type of systems have already been experimentally synthesized and the relevance of the present work leans partially on this fact\cite{PoVaPe2003,SaLiPo2010}. 
Our theoretical analysis suggest the existence of several topological phases within the stripe and the bubble regions of the $H$-$T$ diagrams. 
Although experimental observation and characterization of such intermediate phases is a challenge by itself, the state of the art in magnetic measurement techniques\cite{KrMeZi2015} allows in principle a topological characterization of the phases of the magnetic textures considered here. 

The existence of such exotic phases in magnetic systems is quite interesting because what intuition may suggest is that the temperature would only produce local fluctuations in a way that, either the thermal fluctuations are weak enough to be unable to destroy the magnetic texture or they are just strong enough to destroy the magnetic pattern at once. 
This scenario would be only consistent with the occurrence of a single transition from the low temperature ordered phase to the paramagnetic phase, as is normally predicted by mean-field theories.
Instead, what we find is the occurrence of intermediate topological phases including ``liquid'' states, in which the magnetic textures fluctuate and move as robust structures. 
These states are particularly interesting since the motion of these structures occurs without any mass transportation.

In this work we developed a technique based on the combined use of density functional minimization and the RG equations from the two-dimensional KTHNY melting theory. This approach enables to observe phase transitions not predicted by each theory separately, resulting in rich phase diagrams for ultrathin ferromagnetic films.
Additionally, we established a relation between the strong reentrance of the homogeneous phase and the existence of anomalous or reentrant topological transitions. 
The relation between these two features can be understood in terms of a considerable growth of the modulation length as the external field is increased, producing an anomalous behavior of the elastic constant of the magnetic patterns\cite{PoGoSaBiPeVi2010,MeSt2012}.

\section*{Acknowledgments}

We gratefully acknowledge useful discussions with Mats Wallin and Jack Lidmar.
A.M.C.  acknowledges financial support from Funda\c{c}\~ao de Amparo \`a Pesquisa e Inova\c{c}\~ao do Estado de Santa Catarina (FAPESC).
R.D.M. acknowledges the Swedish Research Council Grants No. 642-2013-7837, 2016-06122, 2018-03659 and G\"{o}ran Gustafsson  Foundation  for  Research  in  Natural  Sciences  and  Medicine and Olle Engkvists Stiftelse.

\bibliographystyle{apsrev4-1}


\end{document}